\documentclass[runningheads]{llncs}

\usepackage[T1]{fontenc}
\usepackage{graphicx}
\usepackage{amsmath}
\usepackage{amsfonts}
\usepackage{xcolor}
\usepackage[normalem]{ulem}  % \sout for strikethrough; normalem preserves \emph
\usepackage{booktabs}
\usepackage{subcaption}
\usepackage{wrapfig}
\usepackage{algpseudocode}
\usepackage{algorithm}
\usepackage{hyperref}
\usepackage{url}
\usepackage{pgfplots}
\pgfplotsset{compat=newest}
\usetikzlibrary{arrows.meta}

\definecolor{UCSDblue}{RGB}{0, 98, 155}
\definecolor{UCSDteal}{RGB}{0, 178, 202}

\newtheorem{assumption}{Assumption}

\ifodd 0

\newcommand{\com}[1]{\textbf{\color{red}(Parinaz: #1)}}
\newcommand{\coml}[1]{\textbf{\color{orange}(Luca: #1)}}
\else

\newcommand{\com}[1]{}
\newcommand{\coml}[1]{}
\fi

\renewcommand{\b}[1]{\boldsymbol{#1}}

\begin{document}

\title{Decision-Focused Learning in\\ Network Interdiction Games}

\author{Luca M. Hartmann\orcidID{0009-0001-5950-4620} \and Parinaz Naghizadeh\orcidID{0000-0002-2277-1709}}

\authorrunning{L. Hartmann and P. Naghizadeh}

\institute{ECE Department, UC San Diego\\
\email{\{lhartmann, parinaz\}@ucsd.edu}}

\maketitle

%%%%%%%%%%%%%%%%%%%%%%%%%
%%%%% Abstract %%%%%%%%%%
%%%%%%%%%%%%%%%%%%%%%%%%%
\begin{abstract}
We study decision-focused learning (DFL) in shortest-path network interdiction (SPNI) games, a Stackelberg game where an interdictor (leader) strengthens the networks' arcs against attacks, while an evader (follower) who is uncertain about costs of attacking network arcs relies on a machine-learned predictor to identify the shortest path. While DFL is highly effective as an end-to-end optimization framework, we show that it faces a fundamental structural failure when employed in this game setting: its training objective admits a broad decision-equivalence class of cost estimators that achieve zero nominal loss yet fail under interdiction, reversing DFL's usual advantage over a naive prediction-focused learning (PFL) approach. To address this, we propose Adversarial DFL (A-DFL), which replaces nominal training samples with interdicted scenarios to collapse the harmful equivalence class. Experiments on synthetic and real-world networks confirm that A-DFL restores DFL's advantage in this game setting, enabling effective end-to-end optimization. 
\keywords{Decision-focused learning \and Network interdiction \and Stackelberg games \and Predict-then-optimize}
\end{abstract}

%%%%%%%%%%%%%%%%%%%%%%%%%
%%%%% Introduction %%%%%%
%%%%%%%%%%%%%%%%%%%%%%%%%
\section{Introduction}
Shortest path network interdiction (SPNI) is a canonical bilevel optimization problem, where a follower (evader) searches for a shortest path through a network, while a leader (interdictor) seeks to fortify the network to maximize the length of the follower's shortest path. Interdictors often represent governmental entities and corporations that defend themselves against criminal organizations. Real-world applications of SPNI include, among others, smuggling prevention~\cite{morton2007}, cyber security~\cite{tanh2017}, and infrastructure protection planning~\cite{jenelius2006}.

Many realistic SPNI scenarios are \emph{asymmetric}, in that the evader lacks precise knowledge of the network's traversal costs. This information asymmetry can be exploited by the interdictor to deploy \emph{asymmetric interdiction strategies}~\cite{bayrak2008asym}, which use the evader's misestimates to lead it along a path that incurs higher costs compared to \emph{symmetric interdiction} actions~\cite{israeli2002benders}, which would be adopted had the evader had full knowledge. 
Given this, it is beneficial for the evader to eliminate or decrease the information asymmetry. 
One way to do so is to leverage auxiliary features that correlate with the true network costs. 
Machine learning (ML) models can exploit these features to predict arc traversal costs, which can then be fed into the evader's shortest-path problem, helping it (partially) mitigate the disadvantages of information asymmetry. 

Such integration of ML models into SPNI games raises questions about how to train prediction models effectively, and how to ensure that they perform well in the strategic setting of SPNI games. 
We address these questions by focusing on an integrated learning and optimization pipeline, which, in the context of combinatorial optimization problems such as shortest paths, is often referred to as ``{decision-focused learning'' (DFL)}~\cite{Mandi2024}. DFL trains the prediction model to minimize \emph{the decision error in the optimization problem}. 
This is in contrast to the classical approach of prediction-focused learning (PFL), which trains the learning model to minimize prediction error, without regard to the eventual use of the predictions in a downstream optimization problem. A line of existing works~\cite{elmachtoub2022spo,Mandi2024} have shown superior performance of DFL over PFL in different end-to-end optimization settings. 

In this paper, we first show that DFL faces a fundamental structural pitfall when deployed in SPNI games: DFL's training objective admits a wide equivalence class of cost estimators that achieve zero training loss, yet differ from the true costs in a way that matters under interdiction. Because of this, despite the attempts by the evador to use DFL strategies to overcome its information asymmetry, the interdictor can still exploit the misalignment to steer the evader onto suboptimal paths. As a result, DFL's usual advantage over the simple, unintegrated baseline of PFL not only collapses but reverses in the game setting; we illustrate this analytically with a toy example in Section~\ref{sec:toy}. 

To address this, we introduce a new training approach, Adversarial DFL (A-DFL), which augments the nominal training samples with interdicted scenarios, collapsing the harmful equivalence class. We introduce two {variants} of A-DFL: RA-DFL (random interdiction selection) and AA-DFL (adversarial worst-case selection). We analyze performance trade-offs between them, and establish their advantages over PFL and DFL, through experiments. Specifically, we validate A-DFL in numerical experiments on grid graphs with varied parameters and a real-world transportation network. 
Results confirm that, while DFL often looses performance compared to PFL in a game setting, {A-DFL} restores DFL's advantage without sacrificing performance. 
We explore the influence of data and model complexity on these findings, and discuss their implications in applications.

To summarize, the main contributions of this paper are as follows:
\begin{enumerate}
    \item We formulate the SPNI problem with a DFL-equipped evader and an interdictor who knows the evader's model, extending classical asymmetric SPNI games to the learning setting (Section~\ref{sec:problem_setup}).
    \item We identify a structural failure of DFL in the game setting: DFL's training objective admits a decision-equivalence class of cost estimators that succeed on the standalone shortest path problem but fail under interdiction, reversing DFL's advantage over PFL in the game setting (Section~\ref{sec:toy}).
    \item We propose {A-DFL, which addresses this failure by replacing nominal training samples with interdicted scenarios (Section~\ref{sec:method}). Our proposed A-DFL has two variants: RA-DFL (random selection) and AA-DFL (adversarial worst-case selection). We analyze their computational and performance trade-offs on synthetic grid graphs and a real-world transportation network (Section~\ref{sec:experiments}).}
\end{enumerate}

For reproducibility, we provide the code of our simulations at: \url{https://github.com/lucahart/DFL-for-Shortest-Path-Network-Interdiction}.

\section{Related Work}
\label{sec:related-work}

SPNI has been studied across a broad range of security domains, including smuggling interdiction~\cite{morton2007}, anti-terrorism operations~\cite{xu2016}, drug trafficking prevention~\cite{magliocca2019}, human trafficking interdiction~\cite{Tezcan2023}, cyber security~\cite{abdallah2020behavioral,tanh2017}, critical infrastructure protection~\cite{jenelius2006}, epidemic control~\cite{assimakopoulos87}, and military and homeland security planning~\cite{brown2009}.\looseness=-1

Several works have considered information asymmetry between interdictors and evaders. One line considers an interdictor with limited knowledge: In~\cite{green2024}, the interdictor cannot observe edge costs, and in~\cite{borrero2016} the interdictor incrementally learns the evader's cost function over repeated interactions. A complementary line considers the opposite asymmetry, where the evader is uncertain but the interdictor retains full or strictly more system information~\cite{Sullivan2014,nguyen2022,azizi2024,mirzaei2021} (the setting most relevant to our work). Closest to our setup is the work of Bayrak and Bailey~\cite{bayrak2008asym}, who propose a bilevel solver for the asymmetric setting in which the interdictor has full information and the evader has limited cost knowledge. Despite the model similarities, unlike this paper, these earlier lines of work do not consider learning and optimization pipelines.

There has been an increasing focus in the ML and OR communities on training of ML models with a downstream optimization problem~\cite{sadana2025,bertsimas2020}. Integrated learning and optimization (ILO) is an end-to-end training paradigm, where the downstream decision error rather than prediction error alone is minimized. The approach dates back to Bengio~\cite{bengio1997} and was later revisited by Donti et al.~\cite{donti2017} and Amos and Kolter~\cite{amos2017optnet} for continuous optimization problems. Combinatorial optimization, in particular, has a range of gradient approximation methods, collectively referred to as \emph{decision-focused learning (DFL)}~\cite{Mandi2024}. First works proposed surrogate losses~\cite{elmachtoub2022spo}, continuous relaxations~\cite{mandi2019}, and smoothing functions~\cite{wilder2018}. More recent work has refined and extended these approaches~\cite{vlastelica2020,mandi2025,wang2026}, with dedicated software libraries supporting their use~\cite{agrawal2019,tang2023pyepo}. DFL has been applied to settings such as last-mile delivery~\cite{chu2023} and ship inspection scheduling~\cite{yan2021}.

Most closely related to our work are Perrault et al.~\cite{perrault2020} and Wang et al.~\cite{wang2020scalable}, who applied DFL to the \emph{leader's} decision pipeline in Stackelberg security games. In contrast, we focus on using DFL for a \emph{follower} with uncertainty. Johnson-Yu et al.~\cite{johnsonyu2023modeling} train an irrational attacker behavioral model to improve defender strategy quality, and \mbox{Żychowski} and Mańdziuk~\cite{zychowski2021learning} learn to predict the behavior of an irrational attacker. 
To the best of our knowledge, no prior work has studied DFL from the evader's perspective in an SPNI game, which is the focus of this paper.\looseness=-1

%%%%%%%%%%%%%%%%%%%%%%%%%
%%%%% Problem Setup %%%%%
%%%%%%%%%%%%%%%%%%%%%%%%%
\section{Problem Setup}
\label{sec:problem_setup}

%%% Stackelberg games %%%
\subsection{Shortest Path Network Interdiction}
We consider the problem of shortest-path network interdiction (SPNI), modeled as a Stackelberg game between an \emph{interdictor} (leader) and an \emph{evader} (follower) on a directed graph $G=(V,E)$ with source $s$ and sink $t$. Each arc $e\in E$ has a \emph{true} base traversal cost $c_e \geq 0$.

The game proceeds as follows. The interdictor acts first by choosing binary interdictions $x_e\in\{0,1\}$ for each arc $e\in E$, under a budget $\sum_{e\in E} x_e \leq B$; if an arc $e$ is interdicted, its traversal cost is increased by an additional known delay $d_e\geq 0$. The evader follows, having observed the vector of interdictions $\b{x}$ selected by the interdictor. Specifically, the evader identifies a shortest path $\b{y}$ (one with the lowest cumulative loss) from source $s$ to target $t$. Knowing this, the interdictor's goal when choosing its interdiction is to maximize the total cost of the path selected by the evader.

Our framework differs from the classic SPNI problem by assuming that the evader has imperfect information about the game: it does not know the base costs $\b{c}$ in $G$, but instead relies on a \emph{learned} cost estimate $\b{\hat{c}}$ when making its decisions. We further assume that both this estimated cost, as well as the true costs $\b{c}$, are known to the interdictor. (We elaborate on the rationale for these assumptions in Section~\ref{sec:assumptions}.)

The problem of finding the Stackelberg equilibrium of this SPNI can therefore be formulated as the following bilevel optimization problem:
\begin{equation}
\label{eq:spni-bilevel}
\max_{\b{x}\in\{0,1\}^{|E|}} \ \ \langle \b{c} + \b{x} \odot \b{d},\, \b{y}\rangle, \quad
\text{s.t.}\ \  \sum_{e\in E} x_e \le B, \quad \b{y} \in \arg\min_{\b{p}\in \mathcal{P}} \ \langle\b{\hat{c}} + \b{x} \odot \b{d},\, \b{p}\rangle,
\end{equation}
where $\mathcal{P}$ denotes the set of $s-t$ path incidence vectors, and $\b{\xi} \odot \b{\phi}$ and $\langle \b{\xi},\, \b{\phi} \rangle$ are the element-wise multiplication and inner product of two vectors $\b{\xi}$ and $\b{\phi}$, respectively.

%%% Learning target %%%
\subsection{Learning and Decision Pipeline}

\begin{figure}
    \vspace{-\baselineskip}
    \centering
    \resizebox{0.9\textwidth}{!}{\begin{tikzpicture}[
    >=stealth,
    neuron/.style={circle, draw=gray!60, fill=gray!25,
                   minimum size=0.4cm, inner sep=0pt, line width=0.5pt},
    gnode/.style={circle, draw=black, fill=white, thick,
                  minimum size=0.5cm, inner sep=0pt, font=\small},
]

%% ---- Neural Network ----
% Input layer
\node[neuron] (in1) at (0,  0.6) {};
\node[neuron] (in2) at (0,  0.0) {};
\node[neuron] (in3) at (0, -0.6) {};

% Hidden layer
\node[neuron] (h1) at (1.2,  .9) {};
\node[neuron] (h2) at (1.2,  0.3) {};
\node[neuron] (h3) at (1.2, -0.3) {};
\node[neuron] (h4) at (1.2, -.9) {};

% Output layer
\node[neuron] (out1) at (2.4,  0.6) {};
\node[neuron] (out2) at (2.4,  0.0) {};
\node[neuron] (out3) at (2.4, -0.6) {};

% Connections: input -> hidden
\foreach \i in {1,2,3}{
    \foreach \j in {1,2,3,4}{
        \draw[gray!65, line width=0.4pt] (in\i) -- (h\j);
    }
}
% Connections: hidden -> output
\foreach \i in {1,2,3,4}{
    \foreach \j in {1,2,3}{
        \draw[gray!65, line width=0.4pt] (h\i) -- (out\j);
    }
}

% f_theta label (centred above hidden layer)
\node[font=\normalsize] at (1.2, 1.3) {$f_{\theta}$};

%% ---- Input / output arrows ----
\draw[->, line width=0.7pt] (-.9, 0) -- (-0.3, 0)
    node[pos=0.3, above, font=\normalsize] {$\boldsymbol{w}$};

\draw[->, line width=0.7pt] (2.65, 0) -- (3.7, 0)
    node[midway, above, font=\normalsize] {$\hat{\boldsymbol{c}}$};

%% ---- Optimization Problem box ----
\node[draw=black, thick, minimum width=2.5cm, minimum height=2.4cm,
      align=center, font=\small, inner sep=6pt] (opt) at (5.1, 0)
{
    \textbf{Optimization}\\[1pt]
    \textbf{Problem}\\[6pt]
    $\displaystyle\min_{\boldsymbol{y}}\; \hat{\boldsymbol{c}}^{\!\top} \boldsymbol{y}$\\[4pt]
    s.t.\ $\boldsymbol{y} \in Y,$
};

% y arrow
\draw[->, line width=0.7pt] (opt.east) -- +(0.8, 0)
    node[midway, above, font=\normalsize] {$\boldsymbol{y}$};

%% ---- Simplified shortest-path graph ----
\begin{scope}[shift={(7.5, 0)}]
    \node[gnode] (s)  at (0,    0)   {$s$};
    \node[gnode] (v1) at (.8,  0.9) {};
    \node[gnode] (v2) at (.8, -0.9) {};
    \node[gnode] (v3) at (1.6,  0)   {};
    \node[gnode] (t)  at (2.6,  0)   {$t$};

    \draw[->, blue!70!black, thick] (s)  -- (v1);
    \draw[->, blue!70!black, thick] (s)  -- (v2);
    \draw[->, blue!70!black, thick] (v1) -- (v3);
    \draw[->, red!80!black,  thick] (v2) -- (v3);
    \draw[->, red!80!black,  thick] (v1) to[bend left=20]  (t);
    \draw[->, blue!70!black, thick] (v2) to[bend right=20] (t);
    \draw[->, blue!70!black, thick] (v3) -- (t);
\end{scope}

%% ---- PFL red box (inner, around NN only) ----
\draw[red!80!black, very thick, rounded corners=5pt]
    (-0.95, -1.4) rectangle (2.85, 1.65);
\node[red!80!black, font=\small\bfseries, anchor=south west,
      align=left, inner sep=3pt, fill=white]
    at (-0.95, 1.65) {Prediction-Focused\\Learning (PFL)};

%% ---- DFL blue box (outer, around everything) ----
\draw[blue!70!black, very thick, rounded corners=7pt]
    (-1.15, -1.6) rectangle (10.85, 2.6);
\node[blue!70!black, font=\small\bfseries, anchor=south west]
    at (-1.15, 2.6) {Decision-Focused Learning (DFL)};

\end{tikzpicture}}
    \caption{The predict then optimize (PO) decision pipeline.}
    \label{fig:PO_framework}
\end{figure}
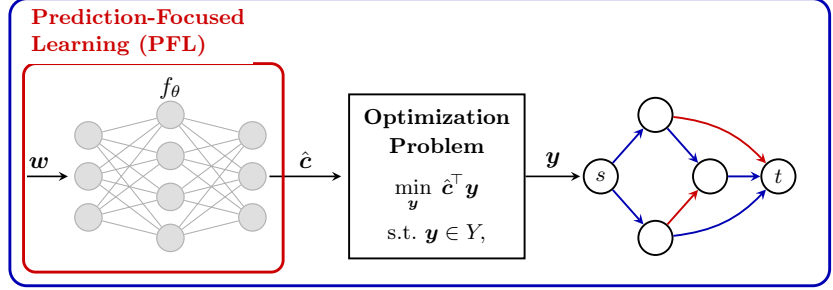

While the true base costs of the edges in the graph are not known to the evader, we assume that it has access to some informative features $\b{w}$, which it can leverage to make edge-costs predictions $\b{\hat c}$. Specifically, we assume that a dataset of features and corresponding true costs is available to the evader. We denote the dataset by $\{(\b{w}^{(k)},\, \b{c}^{(k)})\}_{k=1}^N$, which contains $N$ i.i.d. samples of historical data. The evader trains a (machine learning) model $f_\theta(\b w)$, with parameters $\theta\in\mathbb{R}^p$ based on this training data.
The evader can then use this model to make edge-costs predictions $\b{ \hat{c}}=f_\theta(\b w)$ and subsequently make its attack decision $\b{y}^\ast$ with any classical shortest-path solver. This renders a two-step approach called ``predict-then-optimize'' (PO). Figure~\ref{fig:PO_framework} shows the structure of the PO framework for the example of a shortest path optimization problem.

This work considers two categories of strategies for training the parametric model $f_\theta$: Prediction Focused Learning (PFL) and Decision-Focused Learning (DFL)~\cite{Mandi2024}. PFL refers to the classic training setup, where the model is trained to minimize the prediction error of the costs, without accounting for the downstream prediction task informed by the learned costs. DFL, on the other hand, adopts a holistic view and trains the model to minimize the decision error in the optimization problem. Figure~\ref{fig:PO_framework} contrast these two views within the PO framework. We discuss these learning strategies in detail in Section~\ref{sec:method} and introduce a new mehod called Adversarial DFL (A-DFL), which tailors DFL to the SPNI game setting.

\subsection{Informational Assumptions for the Interdictor and the Evader}
\label{sec:assumptions}

We now formally state the assumptions made in our SPNI setup using motivating examples from \emph{smuggling prevention}.
In this context, the graph nodes are cities, and the edges are paths with exposure log-probabilities (costs) $\b{c}$. The smuggling ring is the evader, which uses reconnaissance to gain information $\b{w}$. The government is the interdictor that chooses binary interdictions $\b{x}$, which increases exposure chances with checkpoints and border patrols by delays $\b{d}$.

\begin{assumption}[game setting]
Each game is played on the same graph structure $G=(V,E)$, but with different costs $\b{c}$ and delays $\b{d}$. \emph{Example.} When {smugglers} want to find a new route over a border, they face new failure log-probabilities $\b{c}$ between new cities and new detection probabilities $\b{d}$ by local law enforcement. \looseness=-1
\end{assumption}

\begin{assumption}[interdictor's information]
The interdictor: (a) has full knowledge of the system, as common in the literature~\cite{bayrak2008asym,Sullivan2014}, and (b) can infer, or reconstruct, the evader's {PFL, DFL, and A-DFL} models through access to the same historic data. \emph{Examples.} (a) The police has knowledge of smuggling success $\b{c}$ and the impact of patrols $\b{x},\,\b{d}$; (b) Governments keep track of smuggling activities on their borders \cite{Tezcan2023}, enabling access to the smuggler's (training) data.
\end{assumption}

\begin{assumption}[evader's information]
The evader: (a) can only infer the costs $\b{c}$ from features $\b{w}$. (b) has accurate data on interdictions of previous systems; {this is relevant for our proposed A-DFL method (both RA-DFL and AA-DFL variants)}; and (c) knows the interdictor's actions $\b{x}$ and delays $\b{d}$. \emph{Examples.}  {Smugglers} (a) don't know the traversal probabilities between different cities when establishing a new route, and can only compare the reconnaissance information with historic routes; (b) can gather observations of the increased failure chances/costs of interdictions on previous attempts; (c) can observe or gather information on new checkpoints and patrols before choosing their route.
\end{assumption}

%%%%%%%%%%%%%%%%%%
%%%%% Methods %%%%
%%%%%%%%%%%%%%%%%%
\section{Methods}
\label{sec:method}

\subsection{Prediction-Focused Learning (PFL)}

PFL refers to the classic training setup, where the model is trained to minimize the prediction error of the costs with classic empirical risk minimization (ERM). PFL trains the model to minimize a loss between the predicted costs $\b{\hat c}$ and the true costs $\b{c}$, without accounting for any downstream prediction task. A common choice for the loss function is the mean squared error (MSE) loss:
\begin{equation}
    \ell_{\mathrm{MSE}}(\b{\hat c};\, \b{c}) \;=\; || \b{\hat c} \;-\; \b{c} ||_2^2~.
\end{equation}

\subsection{Decision-Focused Learning (DFL)}

In contrast to PFL, DFL accounts for the fact that predictions are to be used in a downstream prediction task. To this end, the prediction model $f_\theta$ is trained to minimize regret in the optimization problem's decisions:
\begin{equation} \label{eq:dfl-regret}
    \ell_{\mathrm{DFL}}(\b{\hat c};\, \b{c}) \;=\;  \b{c}^\top \left( \b{y}^\ast(\b{\hat c}) \;-\; \b{y}^\ast(\b{c}) \right)~,
\end{equation}
where $\b{y}^\ast(\b{c})\in\mathcal{Y}$ denotes the solution to the evader's shortest path problem when costs are $\b{c}$. The main challenge in DFL is to compute gradients of the loss $\ell_{\mathrm{DFL}}$ with respect to the predicted costs $\b{\hat c}$. The loss depends on the optimization problem's solution $\b{y}^\ast(\b{\hat c})$, which is typically non-differentiable with respect to $\b{\hat c}$. To address this challenge, various methods have been proposed to compute surrogate gradients through the optimization problem. In our numerical experiments, we will use one such method, SPO$+$~\cite{elmachtoub2022spo}, via the software library~\cite{tang2023pyepo}, to compute a subgradient through a surrogate of the optimization problem. Our findings are not specific to SPO$+$ and generalize to other gradient computation methods.\looseness-1 

The surrogate gradient computed by SPO$+$ is given by
\begin{equation}
    \nabla_{\b{\hat c}} \ell_{\mathrm{DFL}}(\b{\hat c};\, \b{c}) \approx {\b{y}^\ast(\b{c}) - \b{y}^\ast(2\b{\hat c} - \b{c})}~.
\end{equation}
On graphs with cycles, the shifted cost vector $2\hat{\b{c}} - \b{c}$ can produce negative-cost cycles, making the surrogate linear program unbounded. To prevent this, we do the following: When training on a cyclic graph, we clip predicted costs to $\max\{\hat c_i,\, c_i/2+\varepsilon\}$ before evaluating the SPO$+$ loss, and add a penalty term that discourages predictions from falling below this threshold. We detail this safeguard and its effect on gradient propagation in Appendix~\ref{app:spo-fix}.

While DFL is a promising technique for using limited model resources efficiently, it also faces robustness issues. In particular, {DFL can converge to cost estimates that perform well on the training objective but fail under adversarial interdictions, as we demonstrate in Section~\ref{sec:toy}.} Motivated by this, we next describe our proposed method, {A-DFL}, which replaces the nominal training data with interdicted samples to address this failure mode.

\subsection{A-DFL: Adversarial Decision-Focused Learning}

Our proposed method, {Adversarial Decision-Focused Learning (A-DFL)}, 
replaces the evader's nominal training data with interdicted samples, allowing the evader to make robust predictions in games with perturbed cost vectors. Specifically, {A-DFL operates} on two datasets: the existing nominal dataset $\{(\b{w}^{(k)},\, \b{c}^{(k)})\}_{k=1}^N$ of features and true costs, and a newly constructed interdiction dataset \linebreak $\{(\b{d}^{(k,m)},\, \tilde{\b{x}}^{(k,m)})\}_{k=1,\,m=1}^{N,\,N_{\mathrm{intd}}}$ containing $N_{\mathrm{intd}}$ pre-sampled interdiction scenarios per training point. Both the delay vector $\b{d}^{(k,m)}$ and the binary interdiction vector $\tilde{\b{x}}^{(k,m)}$ vary per scenario $m$ and training point $k$. We discuss specific strategies for generating the interdiction scenarios later in this section. 

We make the following assumption on this secondary dataset. 

\begin{assumption}[interdiction dataset]\label{as:intd_dataset}
The interdiction dataset is sampled independently of the nominal dataset and the model's predictions.
\end{assumption}

We define the {A-DFL} decision loss as an average over all $N_{\mathrm{intd}}$ scenarios:
\begin{equation}
\label{eq:rdfl-loss}
    \ell_{\mathrm{A\text{-}DFL}}(\hat{\b{c}}, \b{c})
    \, :=\, \frac{1}{N_{\mathrm{intd}}} \sum_{m=1}^{N_{\mathrm{intd}}}
    \ell_\mathrm{DFL}\!\left(\hat{\b{c}} + \b{d}^{(m)} \odot \tilde{\b{x}}^{(m)},\; \b{c} + \b{d}^{(m)} \odot \tilde{\b{x}}^{(m)}\right).
\end{equation}

It is easy to check that, by construction, $\ell_{\mathrm{A\text{-}DFL}}$ fulfills all properties of a valid decision loss; i.e., $\ell_{\mathrm{A\text{-}DFL}}(\hat{c};c)\ge 0$ and $\ell_{\mathrm{A\text{-}DFL}}(c;c)=0$. 
Further, under Assumption~\ref{as:intd_dataset}, this loss's gradient computation does not require differentiating through the interdiction selection. The training algorithm is summarized in Algorithm~\ref{alg:ADFL}. 

\begin{algorithm}[t]
\caption{Adversarial Decision-Focused Learning (A-DFL)}
\label{alg:ADFL}
\begin{algorithmic}[1]
\Require Learning rate $\eta$, epochs $N_{\mathrm{epochs}}$, nominal dataset $\{(\b{w}^{(k)},\, \b{c}^{(k)})\}_{k=1}^N$, interdiction dataset $\{(\b{d}^{(k,m)},\, \tilde{\b{x}}^{(k,m)})\}_{k=1,\,m=1}^{N,\,N_{\mathrm{intd}}}$, 
\State Initialize model parameters $\theta$
\For{epoch $=1$ to $N_{\mathrm{epochs}}$}
    \For{$k=1$ to $N$}
        \State Predict costs: $\hat{\b{c}}^{(k)} = \hat{f}_\theta(\b{w}^{(k)})$
        \State Compute A-DFL loss: $\ell_{\mathrm{A\text{-}DFL}}^{(k)} = \ldots $ \newline
        \vspace{-.2cm}\begin{flushright} $ \frac{1}{N_{\mathrm{intd}}}\sum_{m=1}^{N_{\mathrm{intd}}} \ell_{DFL}\!\left(\hat{\b{c}}^{(k)} + \b{d}^{(k,m)} \odot \tilde{\b{x}}^{(k,m)},\, \b{c}^{(k)} + \b{d}^{(k,m)} \odot \tilde{\b{x}}^{(k,m)}\right) $ \end{flushright}
        \State Update model parameters: $\theta \leftarrow \theta - \eta \nabla_\theta \ell_{\mathrm{A\text{-}DFL}}^{(k)}$
    \EndFor
\EndFor
\Return Trained model parameters $\theta$
\end{algorithmic}
\end{algorithm}

\vspace{0.2in}
\textbf{A-DFL Variants.} 
We present two variants of A-DFL in this paper. They differ in how they compute the interdiction scenarios $\tilde{\b{x}}^{(k,m)}$ for each training point $k$ and scenario $m$.

The simplest approach is to select $\tilde{\b{x}}^{(k,m)}$ uniformly at random from the set of all feasible interdiction vectors; i.e., those that satisfy the budget constraint. We refer to this method as \emph{Random A-DFL (RA-DFL)}. RA-DFL is computationally cheap, fulfills Assumption~\ref{as:intd_dataset}, and provides diverse interdiction scenarios.\looseness=-1

An alternative approach, which we call \emph{Adversarial A-DFL (AA-DFL)}, selects $\tilde{\b{x}}^{(k,m)}$ as the solution to the worst-case bilevel optimization problem using the true costs $\b{c}^{(k)}$; i.e., the interdiction that a symmetric interdictor~\cite{israeli2002benders} would choose. Because the selection depends only on the true costs and not on the model's predictions $\hat{\b{c}}^{(k)}$, Assumption~\ref{as:intd_dataset} is satisfied and gradient computation remains tractable. 

AA-DFL is computationally expensive, since it requires solving an NP-hard bilevel optimization problem per training point and interdiction scenario in a preprocessing stage. We notice no significant improvement over RA-DFL when training it on small graphs. This can be expected, as a relatively large number of interdictions $B$ relative to $|\mathcal{E}|$ is likely to cover all relevant interdiction scenarios.

In contrast, on larger graphs with a small budget $B$ relative to $|\mathcal{E}|$, uniform random sampling is unlikely to select all relevant arcs. Consequently, RA-DFL behaves more similar to DFL, whereas the more focused adversarial selection can be more effective. We investigate the trade-offs between RA-DFL and AA-DFL empirically in Section~\ref{sec:experiments}.

%%%%%%%%%%%%%%%%%%%%%%%%%
%%%%% Toy Example %%%%%%%
%%%%%%%%%%%%%%%%%%%%%%%%%
\section{Toy Example: Why DFL Fails Under Interdiction}
\label{sec:toy}

\subsection{Problem Setup}

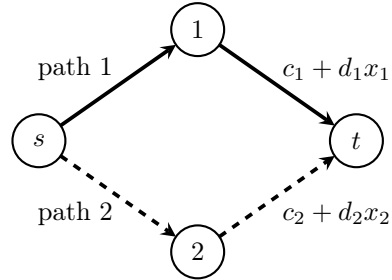
\begin{wrapfigure}[12]{r}{0.42\textwidth}
    \vspace{-0.2in}
    \centering
    \begin{tikzpicture}[
    x=0.42cm, y=0.42cm,
    node/.style={circle, draw=black,
                 thick, minimum size=0.7cm, inner sep=2pt,
                 font=\normalsize},
    >=stealth, thick,
    every node/.append style={outer sep=0pt}
  ]
  \node[node] (s) at (0,    0)   {$s$};
  \node[node] (1) at (5,  3.5)   {$1$};
  \node[node] (2) at (5, -3.5)   {$2$};
  \node[node] (t) at (10,   0)   {$t$};
  \draw[line width=0.5mm, ->] (s) -- node[above left,  font=\normalsize, inner sep=2pt]{path 1} (1);
  \draw[line width=0.5mm, ->] (1) -- node[above right, font=\normalsize, inner sep=2pt]{$c_1 + d_1 x_1$} (t);
  \draw[line width=0.5mm, ->, dashed] (s) -- node[below left,  font=\normalsize, inner sep=2pt]{path 2} (2);
  \draw[line width=0.5mm, ->, dashed] (2) -- node[below right, font=\normalsize, inner sep=2pt]{$c_2 + d_2 x_2$} (t);
\end{tikzpicture}
    \vspace{-0.2in}
    \caption{The toy graph.}
    \label{fig:toy_graph}
\end{wrapfigure}

To illustrate the pitfalls of DFL in game settings and motivate A-DFL, we use a toy SPNI game on a graph with two paths. The toy graph is depicted in Figure~\ref{fig:toy_graph}, and for the remainder, we will refer to the paths as \emph{path 1} (s$\rightarrow$ 1 $\rightarrow$ t; solid lines) and \emph{path 2} (s $\rightarrow$ 2 $\rightarrow $ t; dotted lines), with costs $c_1$ and $c_2$, respectively. 

\begin{figure*}[t]
    \centering
    % ---- Column 1: true costs ----
    \begin{minipage}[t]{0.31\textwidth}
        \begin{subfigure}[b]{\linewidth}
            \centering
            \resizebox{\linewidth}{!}{\begin{tikzpicture}
    \begin{axis}[
        width=7cm,
        height=7cm,
        xlabel={$w$},
        ylabel={$c_i$},
        xmin=-6, xmax=6,
        ymin=-6.6, ymax=6.6,
        axis lines=center,
        grid=major,
        grid style={dashed, gray!40},
        xtick={-6,-3,0,3,6},
        ytick={-6,-3,0,3,6},
        tick label style={font=\normalsize},
        label style={font=\normalsize},
        legend pos=north west,
        legend style={font=\normalsize, draw=none},
    ]
        \addplot[domain=-6:0, samples=2, ultra thick, forget plot] {x};
        \addplot[domain=0:6, samples=2, thick] {x};
        \addlegendentry{$i=1$}
        \addplot[domain=-6:0, samples=2, thick, dashed] {-x};
        \addlegendentry{$i=2$}
        \addplot[domain=0:6, samples=2, ultra thick, dashed, forget plot] {-x};
    \end{axis}
\end{tikzpicture}}
            \caption{True, uninterdicted.}
            \label{fig:toy-unintd}
        \end{subfigure}

        \medskip

        \begin{subfigure}[b]{\linewidth}
            \centering
            \resizebox{\linewidth}{!}{\begin{tikzpicture}
    \begin{axis}[
        width=7cm,
        height=7cm,
        xlabel={$w$},
        ylabel={$c_i + d_i x_i$},
        xmin=-6, xmax=6,
        ymin=-6.6, ymax=6.6,
        axis lines=center,
        grid=major,
        grid style={dashed, gray!40},
        xtick={-6,-3,0,3,6},
        ytick={-6,-3,0,3,6},
        tick label style={font=\normalsize},
        label style={font=\normalsize},
        legend pos=north west,
        legend style={font=\normalsize, draw=none},
    ]
        % True costs, path 1
        \addplot[domain=-6:-1.5, samples=2, ultra thick, forget plot] {x + 3};
        \addplot[domain=-1.5:0, samples=2, thick] {x + 3};
        \addplot[domain=0:1.5, samples=2, ultra thick, forget plot] {x};
        \addplot[domain=1.5:6, samples=2, thick, forget plot] {x};
        \addlegendentry{$i=1$}

        % True costs, path 2
        \addplot[domain=-6:-1.5, samples=2, thick, dashed] {-x};
        \addplot[domain=-1.5:0, samples=2, ultra thick, dashed, forget plot] {-x};
        \addplot[domain=0:1.5, samples=2, thick, dashed] {-x + 3};
        \addplot[domain=1.5:6, samples=2, ultra thick, dashed] {-x + 3};
        \addlegendentry{$i=2$}
    \end{axis}
\end{tikzpicture}}
            \caption{True, interdicted.}
            \label{fig:toy-intd}
        \end{subfigure}
    \end{minipage}
    \hfill
    % ---- Column 2: DFL ----
    \begin{minipage}[t]{0.31\textwidth}
        \begin{subfigure}[b]{\linewidth}
            \centering
            \resizebox{\linewidth}{!}{\begin{tikzpicture}
    \begin{axis}[
        width=7cm,
        height=7cm,
        xlabel={$w$},
        ylabel={$c_i$},
        xmin=-6, xmax=6,
        ymin=-6.6, ymax=6.6,
        axis lines=center,
        grid=major,
        grid style={dashed, gray!40},
        xtick={-6,-3,0,3,6},
        ytick={-6,-3,0,3,6},
        tick label style={font=\normalsize},
        label style={font=\normalsize},
        legend pos=north west,
        legend style={font=\normalsize, draw=none},
    ]
        % True costs (in black)
        \addplot[domain=-6:0, samples=2, ultra thick, forget plot] {x};
        \addplot[domain=0:6, samples=2, thick, forget plot] {x};
        \addlegendentry{$i=1$}
        \addplot[domain=-6:0, samples=2, thick, dashed, forget plot] {-x};
        \addlegendentry{$i=2$}
        \addplot[domain=0:6, samples=2, ultra thick, dashed, forget plot] {-x};

        % Predicted costs (in cyan)
        \addplot[domain=-6:0, samples=2, ultra thick, color=cyan, forget plot] {0.6*x};
        \addplot[domain=0:6, samples=2, thick, color=cyan] {0.6*x};
        \addlegendentry{$i=1$}
        \addplot[domain=-6:0, samples=2, thick, color=cyan, dashed] {-0.6*x};
        \addplot[domain=0:6, samples=2, ultra thick, color=cyan, dashed, forget plot] {-0.6*x};
        \addlegendentry{$i=2$}
    \end{axis}
\end{tikzpicture}}
            \caption{DFL, uninterdicted.}
            \label{fig:toy-dfl-unintd}
        \end{subfigure}

        \medskip

        \begin{subfigure}[b]{\linewidth}
            \centering
            \resizebox{\linewidth}{!}{\begin{tikzpicture}
    \begin{axis}[
        width=7cm,
        height=7cm,
        xlabel={$w$},
        ylabel={$c_i + d_i x_i$},
        xmin=-6, xmax=6,
        ymin=-6.6, ymax=6.6,
        axis lines=center,
        grid=major,
        grid style={dashed, gray!40},
        xtick={-6,-3,0,2,3,6},
        ytick={-6,-3,0,3,6},
        tick label style={font=\normalsize},
        label style={font=\normalsize},
        legend pos=north west,
        legend style={font=\normalsize, draw=none},
    ]
        % Red marked areas
        \fill[red, fill opacity=0.2, draw=none] (axis cs:1.5,-6.6) rectangle (axis cs:2.5,6.6);
        \fill[red, fill opacity=0.2, draw=none] (axis cs:-2.5,-6.6) rectangle (axis cs:-1.5,6.6);

        % True costs, path 1
        \addplot[domain=-6:-1.5, samples=2, ultra thick, forget plot] {x + 3};
        \addplot[domain=-1.5:0, samples=2, thick, forget plot] {x + 3};
        \addplot[domain=0:1.5, samples=2, ultra thick, forget plot] {x};
        \addplot[domain=1.5:6, samples=2, thick, forget plot] {x};
        \addlegendentry{$i=1$}

        % True costs, path 2
        \addplot[domain=-6:-1.5, samples=2, thick, dashed, forget plot] {-x};
        \addplot[domain=-1.5:0, samples=2, ultra thick, dashed, forget plot] {-x};
        \addplot[domain=0:1.5, samples=2, thick, dashed, forget plot] {-x + 3};
        \addplot[domain=1.5:6, samples=2, ultra thick, dashed, forget plot] {-x + 3};
        \addlegendentry{$i=2$}

        % Estimated costs, path 1
        \addplot[domain=-6:-2.5, samples=2, ultra thick, color=cyan, forget plot] {0.6*x + 3};
        \addplot[domain=-2.5:0, samples=2, thick, color=cyan, forget plot] {0.6*x + 3};
        \addplot[domain=0:2.5, samples=2, ultra thick, color=cyan, forget plot] {0.6*x};
        \addplot[domain=2.5:6, samples=2, thick, color=cyan] {0.6*x};
        \addlegendentry{$i=1$}

        % Estimated costs, path 2
        \addplot[domain=-6:-2.5, samples=2, thick, color=cyan, dashed, forget plot] {-0.6*x};
        \addplot[domain=-2.5:0, samples=2, ultra thick, color=cyan, dashed, forget plot] {-0.6*x};
        \addplot[domain=0:2.5, samples=2, thick, color=cyan, dashed] {-0.6*x + 3};
        \addplot[domain=2.5:6, samples=2, ultra thick, color=cyan, dashed, forget plot] {-0.6*x + 3};
        \addlegendentry{$i=2$}
    \end{axis}
\end{tikzpicture}}
            \caption{DFL, interdicted.}
            \label{fig:toy-dfl-intd}
        \end{subfigure}
    \end{minipage}
    \hfill
    % ---- Column 3: A-DFL ----
    \begin{minipage}[t]{0.31\textwidth}
        \begin{subfigure}[b]{\linewidth}
            \centering
            \resizebox{\linewidth}{!}{\begin{tikzpicture}
    \begin{axis}[
        width=7cm,
        height=7cm,
        xlabel={$w$},
        ylabel={$c_i$},
        xmin=-6, xmax=6,
        ymin=-6.6, ymax=6.6,
        axis lines=center,
        grid=major,
        grid style={dashed, gray!40},
        xtick={-6,-3,0,3,6},
        ytick={-6,-3,0,3,6},
        tick label style={font=\normalsize},
        label style={font=\normalsize},
        legend pos=north west,
        legend style={font=\normalsize, draw=none},
    ]
        % True costs
        \addplot[domain=-6:0, samples=2, ultra thick, forget plot] {x};
        \addplot[domain=0:6, samples=2, thick, forget plot] {x};
        \addplot[domain=-6:0, samples=2, thick, dashed, forget plot] {-x};
        \addplot[domain=0:6, samples=2, ultra thick, dashed, forget plot] {-x};
        % Estimated costs
        \addplot[domain=-6:0, samples=2, ultra thick, color=orange, forget plot] {0.95*x};
        \addplot[domain=0:6, samples=2, thick, color=orange] {0.95*x};
        \addlegendentry{$i=1$}
        \addplot[domain=-6:0, samples=2, thick, color=orange, dashed] {-0.95*x};
        \addplot[domain=0:6, samples=2, ultra thick, color=orange, dashed, forget plot] {-0.95*x};
        \addlegendentry{$i=2$}
    \end{axis}
\end{tikzpicture}}
            \caption{A-DFL, uninterdicted.}
            \label{fig:toy-adfl-unintd}
        \end{subfigure}

        \medskip

        \begin{subfigure}[b]{\linewidth}
            \centering
            \resizebox{\linewidth}{!}{\begin{tikzpicture}
    \begin{axis}[
        width=7cm,
        height=7cm,
        xlabel={$w$},
        ylabel={$c_i + d_i x_i$},
        xmin=-6, xmax=6,
        ymin=-6.6, ymax=6.6,
        axis lines=center,
        grid=major,
        grid style={dashed, gray!40},
        xtick={-6,-3,0,3,6},
        ytick={-6,-3,0,3,6},
        tick label style={font=\normalsize},
        label style={font=\normalsize},
        legend pos=north west,
        legend style={font=\normalsize, draw=none},
    ]
        % True costs
        \addplot[domain=-6:-1.5, samples=2, ultra thick, forget plot] {x + 3};
        \addplot[domain=-1.5:0, samples=2, thick, forget plot] {x + 3};
        \addplot[domain=0:1.5, samples=2, ultra thick, forget plot] {x};
        \addplot[domain=1.5:6, samples=2, thick, forget plot] {x};
        \addplot[domain=-6:-1.5, samples=2, thick, dashed, forget plot] {-x};
        \addplot[domain=-1.5:0, samples=2, ultra thick, dashed, forget plot] {-x};
        \addplot[domain=0:1.5, samples=2, thick, dashed, forget plot] {-x + 3};
        \addplot[domain=1.5:6, samples=2, ultra thick, dashed, forget plot] {-x + 3};
        % Estimated costs
        \addplot[domain=-6:-1.5789, samples=2, ultra thick, color=orange, forget plot] {0.95*x + 3};
        \addplot[domain=-1.5789:0, samples=2, thick, color=orange, forget plot] {0.95*x + 3};
        \addplot[domain=0:1.5789, samples=2, ultra thick, color=orange, forget plot] {0.95*x};
        \addplot[domain=1.5789:6, samples=2, thick, color=orange] {0.95*x};
        \addlegendentry{$i=1$}
        \addplot[domain=-6:-1.5789, samples=2, thick, color=orange, dashed, forget plot] {-0.95*x};
        \addplot[domain=-1.5789:0, samples=2, ultra thick, color=orange, dashed, forget plot] {-0.95*x};
        \addplot[domain=0:1.5789, samples=2, thick, color=orange, dashed] {-0.95*x + 3};
        \addplot[domain=1.5789:6, samples=2, ultra thick, color=orange, dashed, forget plot] {-0.95*x + 3};
        \addlegendentry{$i=2$}
    \end{axis}
\end{tikzpicture}}
            \caption{A-DFL, interdicted.}
            \label{fig:toy-adfl-intd}
        \end{subfigure}
    \end{minipage}

    \caption{True cost functions (black) and estimated cost functions under DFL (blue) and A-DFL (orange). The $x$-axis corresponds to the feature, and the $y$-axis is the (true or estimated) cost function for path 1 (solid lines) and path 2 (dotted lines). The optimal path choice at each feature $\omega$ is the one with the lower (true or estimated) cost on the $y$-axis; thick (solid/dotted) lines are used to highlight these. Accordingly, the red shaded regions in plot (d) show a disagreement between the optimal solution and that selected by DFL under interdiction.}
    \label{fig:toy_example}
\end{figure*}
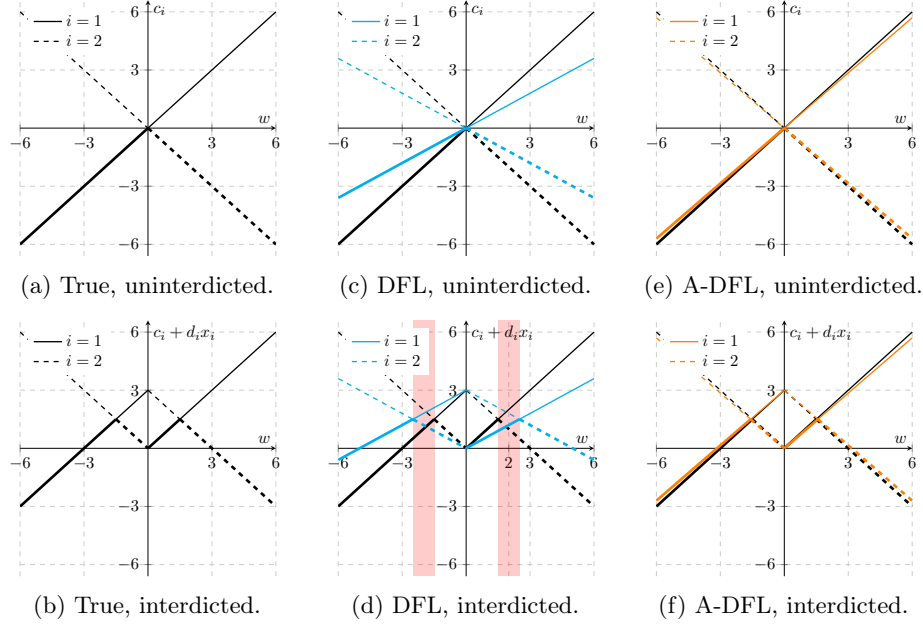

We choose a linear mapping $f:\mathbb{R}^2 \to \mathbb{R}^2$ from features $\b w \in \mathbb{R}^2$ to costs $\b c \in \mathbb{R}^2$:
\begin{equation}
\label{eq:feature-cost-mapping}
    \begin{bmatrix}
        c_1 \\
        c_2
    \end{bmatrix} =
    f(w)=
    \begin{bmatrix}
        w \\
        -w
    \end{bmatrix}
\end{equation}
The function $f$ is plotted in Figure~\ref{fig:toy-unintd}. 

\paragraph{The evader's problem.} We assume the evader uses an affine prediction model of the form $\hat{\b c} = \hat f(w) = \b a w + \b b,$ where $\b a \in \mathbb R^2$ and $\b b\in\mathbb R^2 $ are learned parameters. The model is trained using either DFL or A-DFL on 5000 samples, 500 epochs. We always initialize the model with $a_1 = -a_2 = 0.1$ and $\b b = 0$ for visualization purposes and without loss of generality. For more details see Appendix~\ref{app:toy}. The high number of training samples and epochs is chosen to ensure convergence the optimal solution. We did not observe significant changes in the results when increasing the number of samples or epochs.

\paragraph{The interdictor's problem.} We assume a budget of $B=1$, so that the interdictor can only choose to interdict one path. For simplicity and without loss of generality, we assume that the evader's cost estimates are always good enough so that the asymmetric interdictor is equal to the symmetric interdictor. Consequently, the interdictor always chooses to interdict the path with the smaller true cost. 
We discuss this assumption and its implications in Appendix~\ref{app:toy}.

\paragraph{The optimal (full information) evader decisions.} Recall that the evader's decision $\b y^\ast$ is given by the solution to the shortest path problem:
\begin{align}
\label{eq:toy-shortest-path}
    \b y^\ast = &\arg \min_y (c_1 + d_1 x_1) y_1 + (c_2 + d_2 x_2) y_2,\notag \\
    \text{s.t. }&~~ y_1 + y_2 \geq 1, ~~ y_1, y_2 \in \{0,1\},
\end{align}
where $\b y \in \{0,1\}^2$ is the one-hot decision variable for the chosen shortest path, $\b x \in \{0,1\}^2$ is the interdictor's decision vector, and $\b d = [d_1,\, d_2]^\top = [3,\, 3]^\top$ is the delay vector. In the toy case, this simplifies to choosing the path with the smaller cost $c_1 + d_1 x_1$ or $c_2 + d_2 x_2$; as such, in Figure~\ref{fig:toy_example}, the evader always chooses the path associated to the lowest true/estimated cost on the $y$-axis. The optimal solutions are marked with bold lines in Figures~\ref{fig:toy-unintd} and~\ref{fig:toy-intd}. 

\vspace{-0.07in}
\subsection{Analysis}

\vspace{-0.03in}
We now discuss the results of DFL and A-DFL, depicted in Figures~\ref{fig:toy-dfl-unintd} - \ref{fig:toy-adfl-intd}. 

\vspace{0.05in}
\textbf{DFL before interdiction.} Figure~\ref{fig:toy-dfl-unintd} shows the DFL-trained model and its decision. Although the estimated cost function differs from the true function, the cost \emph{ordering} between path 1 and path 2 is preserved for all feature values $w$, so the optimizer obtained from~\eqref{eq:toy-shortest-path} based on these estimated costs always selects the correct shortest path. Consequently, the DFL regret~\eqref{eq:dfl-regret} is zero. Thus, the DFL gradient vanishes and the incorrect slopes are not penalized during training. This also holds for the region of slopes $\b a$ around the given solution that keep the same ordering of estimated path costs. 

Formally, for a linear model $\hat f(w) = \b a w + \b b$, we define the \emph{decision-equivalence class} $\mathcal{F}_{DFL}$ as:
\begin{equation}
    \label{eq:dfl-set}
    \mathcal F_{DFL} = \{f : \mathbb R^2 \rightarrow \mathbb R^2,\, f(w) = \b a w + \b b\, |\, b_1 = b_2,\, a_1 > a_2 \}.
\end{equation}
We obtain a zero DFL loss, $\ell_{DFL} = \b c^\top (\b y^\ast(\hat f(w)) - \b y^\ast (f(w))) = 0$, if and only if the estimated function is part of the decision-equivalence class $\hat f(w) \in \mathcal F_{DFL}$. This leads to a flat cost region, resulting in zero gradients $\nabla_{\b a} \ell_{DFL} = 0$. We refer the reader to Appendix~\ref{app:toy} for more details on the derivation of $\mathcal{F}_{DFL}$ and its implications for DFL training.

The estimated model (blue) in Figure~\ref{fig:toy-dfl-unintd} lies in $\mathcal{F}_{DFL}$.
This {supports our assumption} that the incorrect slopes are a consequence of this decision-equivalence class rather than insufficient training.\footnote{{Training itself uses the SPO$+$ surrogate rather than the exact regret~\eqref{eq:dfl-regret}; the SPO$+$ zero-loss set for this toy example is the strict subset $\{a_1 - a_2 > 1,\, b_1 = b_2\} \subset \mathcal F_{DFL}$, 
so the trained slope landing in $\mathcal F_{DFL}$ is in fact predicted by, not merely consistent with, training on SPO$+$.}}

\vspace{0.1in}
\textbf{DFL after interdiction.} Next, we embed the DFL trained model into the SPNI game. Figure~\ref{fig:toy-dfl-intd} shows the true and estimated costs after interdiction. Observe that there is now an 
incorrect ordering of the estimated costs and true costs of the path in an interval $\mathcal C_{err}$ (marked in red in the figure), which leads to incorrect decisions by DFL. For instance, at $w=2$, the DFL-estimated model (blue) believes that path 1 is cheaper than path 2. This leads to the incorrect decision of choosing path 1, whereas the true model would choose path 2. We conclude that, while DFL succeeds in the uninterdicted scenario, Figure~\ref{fig:toy-dfl-intd} shows that it fails, across a range of inputs, when embedded in an SPNI game. 

To explain why this happens, note that for correct decisions after interdiction, the estimated model must also preserve the correct cost ordering under the shifted costs $\hat{\b c} + \b d \b x$. In the toy example, this requires the \emph{relative slope} $a_1 - a_2$ to equal the true value of $2$; i.e., the angle between the path costs of the estimates must be equal to the true angle. 
We formalize this requirement by defining a tighter decision equivalence class, $\mathcal F_{A\text{-}DFL} \subset \mathcal F_{DFL}$, as follows:
\begin{equation}
    \label{eq:adfl-set}
    \mathcal F_{A\text{-}DFL} = \{f : \mathbb R \rightarrow \mathbb R,\, f(w) = \b a w + \b b\, |\, b_1 = b_2,\, a_1 - a_2 = 2\},
\end{equation}
Every $\hat f \in \mathcal F_{A\text{-}DFL}$ achieves correct decisions both before and after interdiction. {We relegate the proof to Appendix~\ref{app:toy}}. The DFL-trained model in Figure~\ref{fig:toy-dfl-intd} lies in $\mathcal F_{DFL} \setminus \mathcal F_{A\text{-}DFL}$: its relative slope is unequal to $2$, so the estimated cost ordering breaks down once delays are applied. 

\vspace{0.1in}
\textbf{A-DFL.} First, Figure~\ref{fig:toy-adfl-unintd} confirms that A-DFL makes the optimal decisions in the absence of any interdiction. Furthermore, after interdiction, unlike the DFL loss, the A-DFL loss is nonzero:
\begin{equation*}
    \ell_{A\text{-}DFL}(\hat{\b c};\b c) = (\b c + \b d \odot \b x)^\top \bigl(\b y^\ast(\hat{\b c} + \b d \odot \b x) - \b y^\ast(\b c + \b d \odot \b x)\bigr) \neq 0,
\end{equation*}
since the evader makes the wrong path choice on interdicted samples whenever $\hat c \in \mathcal C_{err}$. {This coincides with our findings in~\eqref{eq:adfl-set} that $\ell_{\mathrm{A\text{-}DFL}} = 0$ iff $\hat f \in \mathcal F_{\mathrm{A\text{-}DFL}}$. Our experiments in Figure~\ref{fig:toy-adfl-intd} confirm that A-DFL converges to a model in the tighter class $\mathcal F_{A\text{-}DFL}$ that achieves correct decisions with and without interdiction.}

We note that $\mathcal{F}_{A\text{-}DFL}$ still contains infinitely many models: fixing the relative slope $a_1 - a_2 = 2$ {leaves two residual degrees of freedom, the common intercept $b_1 = b_2$ and the common slope level (e.g., $a_1 + a_2$, with $a_1 - a_2 = 2$).} A-DFL does not fully identify the true cost function either, only the decision-relevant direction.
We conjecture that when the prediction model is not expressive enough to approximate the true feature-cost mapping (as is the case in our toy example), A-DFL is using its degrees of freedom more efficiently than DFL because of its use of the secondary (either randomly or synthetically-generated) interdiction dataset. 
We leave a formal investigation of this conjecture to future work. 

%%%%%%%%%%%%%%%%%%%%%%%%%%
%%%%% Experiments %%%%%%%%
%%%%%%%%%%%%%%%%%%%%%%%%%%
\section{Experiments}
\label{sec:experiments}

%\begin{sloppypar}
We now present the setup and results from our numerical experiments. For reproducibility, we provide the code of our simulations at: \url{https://github.com/lucahart/DFL-for-Shortest-Path-Network-Interdiction}.

\subsection{Experimental Setup}

\begin{figure}
    \centering
    \begin{subfigure}[b]{0.48\textwidth}
        \centering
        \includegraphics[width=0.75\textwidth]{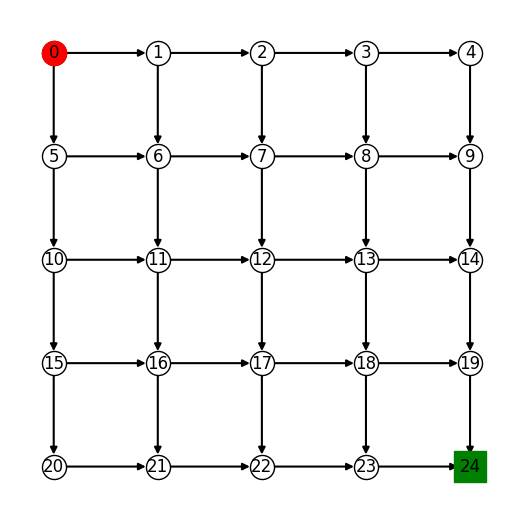}
        \caption{Grid graph topology.}
        \label{fig:5x5-grid}
    \end{subfigure}\hfill
    \begin{subfigure}[b]{0.48\textwidth}
        \centering
        \includegraphics[width=\textwidth]{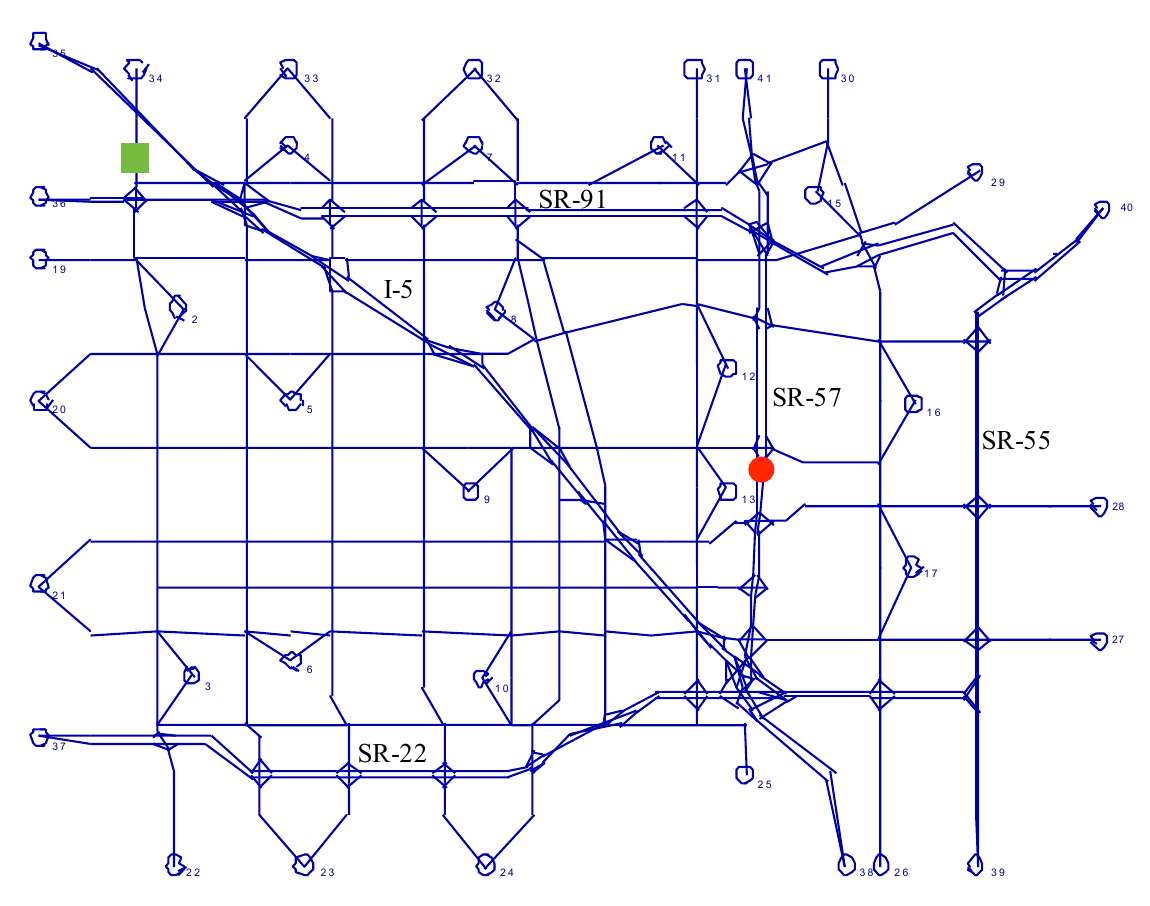}
        \caption{Real-world graph topology.}
        \label{fig:real-world-graph}
    \end{subfigure}
    \caption{Graph topologies used in the experiments. The source node is marked with a red dot, and the target node is marked with a green rectangle.}
    \label{fig:graph-topologies}
\end{figure}

\subsubsection{Training Data Generation.}

We generate synthetic training data similar to~\cite{elmachtoub2022spo}. The data generation function samples $p$ input features from a normal distribution $x \sim N(0,I_p)$. The generator function then computes the costs:
\begin{equation}
    \label{eq:data-dist}
    c_{ij} = \left[  \left( \frac{1}{\sqrt{p}} \b B x_i)_j + 3 \right)^{deg} + 1 \right] \epsilon_i^j.
\end{equation}
Here, $j = 1,\ldots,r$ are the costs in the cost vector $\b c \in \mathbb R^r$ and $i = 1,\ldots, N$ are the data samples. The parameter $deg$ defines the degree of the function; throughout the experiments, we vary the degree of this generator function to show the effect of increasing complexity of the mapping from costs to features. $\epsilon$ is a multiplicative noise drawn from a uniform distribution on the interval $[1-\bar \epsilon, 1 + \bar \epsilon]$. The matrix $B \in \mathbb R ^{r \times p}$ is sampled from a Bernoulli random variable that is equal to 1 with probability $0.5$. The delay vectors $\b{d}^{(k,m)}$ in the interdiction dataset are generated using the same distribution~\eqref{eq:data-dist}, with independently drawn features and noise realizations, so that delays have the same scale and structure as the nominal costs. The features of the delays are then discarded.

\subsubsection{Synthetic grid graph.} Our first evaluation is on a directed grid graph with $m \times n = 5\times 5$ nodes. We denote $m$ as the number of rows and $n$ as the number of columns. Figure~\ref{fig:5x5-grid} displays the grid graph topology. It is a standard in the DFL literature~\cite{elmachtoub2022spo,Mandi2024} and has also been used in the SPNI literature~\cite{song2016}. The evader's task is to find the shortest path from the top-left node (denoted as $0$) to the bottom-right node (denoted as $mn - 1$). This setup provides the evader with $\binom{m + n - 2}{n - 1} = 70$ paths that all have a length of $m + n - 2$ edges. Given i.i.d. sampled edge weights, this creates a challenging scenario with many paths of similar length. Given the small graph size, we use affine functions for the prediction model. Depending on the feature-cost mapping's complexity, this allows us to model cases where the model is sufficiently powerful to learn the true mapping, as well as cases where the model is not powerful enough. 

\subsubsection{Real-world graph.}  We also confirm our findings on a real-world graph, namely the Anaheim graph~\cite{transportationnetworks}, displayed in Figure~\ref{fig:real-world-graph}. It is a directed graph with 416 nodes and 914 edges. All simulations consider the path from node 108 to 410, which has a shortest path of 16 edges length without weights. We use the same data generation process as in the grid simulations, but increase the feature-cost mapping complexity to complement the increased path complexity in the real-world graph. This new setup is too complex for a simple affine predictor. Thus, the parametric learning model is a neural network with one hidden layer with 8 neurons and a positive sigmoid activation function to avoid negative costs. Again, this allows us to set the feature-cost mapping to be simple enough and too complex for the parametric model. 

\subsubsection{Training and testing procedures.} 
A single simulation trains {four} models using gradients generated from PFL, DFL, {RA-DFL, and AA-DFL}. The models are then tested on networks with test values sampled from the same distribution as the training samples. All simulation parameters are displayed in Table~\ref{tab:sim-params}. 

\begin{table}
    \centering
    \caption{Simulation parameters for the grid and real-world graph experiments.}
    \begin{tabular}{ccc}
        \toprule
        Parameter Name & Parameter value (grid) & Parameter value (real-world)\\
        \midrule
        num. training data & $N = 500$ & $N = 500$ \\
        num. testing data & $N_{test} = 250$ & $N_{test} = 250$ \\
        budget & $ B= 5 $ & $ B= 5 $ \\
        grid size & $(m,n) = (5, 5)$ & N/A \\
        {random} intd. scenarios & $3$ & $3$ \\
        number of simulations & $N_{sim} = 5$ & $N_{sim} = 5$ \\
        num. features & $p = 5$  & $p = 5$ \\
        num. epochs PFL & 800 & 800\\
        num. epochs DFL & 600 & 400\\
        PFL learning rate & $l_{PFL} = 1 \times 10^{-2}$ & $l_{PFL} = 5 \times 10^{-2}$\\
        DFL learning rate & $l_{DFL} = 1 \times 10^{-2}$ & $l_{DFL} = 1 \times 10^{-2}$\\
        gen. function degree & $deg \in \{{3}, 8\}$ & $deg \in \{10, 12\}$ \\
        noise width & $\bar \epsilon = 0.5$ & $\bar \epsilon = 0.5$\\
        \bottomrule
    \end{tabular}
    \label{tab:sim-params}
\end{table}

\subsection{Evaluation Metrics}

The prediction models are evaluated by calculating a simulation regret $R_{sim}$ that uses the sample regret $R$. Thus, we present the sample regret and simulation regret consecutively.

For each test sample we run a prediction of the costs $\hat c_i$ from the feature $w_i$. Then, we solve the interdictor's optimization problem~\eqref{eq:spni-bilevel} given $\hat c_i$ and compute the sample regret
\begin{equation*}
    {R}(\hat{\b c} + \b d \odot \b x; \b c + \b d \odot \b x) = \ell_{DFL}(\hat{\b c} + \b d \odot \b x; \b c + \b d \odot \b x).
\end{equation*}

To compute the simulation regret, we run $N_{sim}$ simulations with the same parameters. In each simulation, new training and testing data is generated from a different seed. We then compute the normalized regret over the entire test set of a simulation: 
\begin{equation*}
    {R}_{sim} = \frac{\sum_{i=1}^{N_{test}} {R}(\hat{ \b c}_i + \b d_i \odot \b x_i; \b{ c}_i + \b d_i \odot \b x_i)}{\sum_{i=1}^{N_{test}} (\b c_i + \b d_i \odot \b x_i)^\top \b y^\ast (\b c_i + \b d_i \odot \b x_i)}.
\end{equation*}
This loss can also be interpreted as the percentage increase in cost of the evader's path over the oracle's optimal path. Thus, we will use the unit $[\%]$ for the simulation regret.

When evaluating the game setting, we use new values for the interdictions $\b d$ at every test sample. This procedure is crucial to show how {RA-DFL (and AA-DFL)} generalize. For reference, we also compute the simulation regret for the uninterdicted problem (shortest path problem without embedding into a game) by choosing $\b x = \b 0$.

\subsection{Main Results}

\begin{figure}[t]
    \begin{subfigure}[b]{0.49\textwidth}
        \centering
        \includegraphics[width=\textwidth]{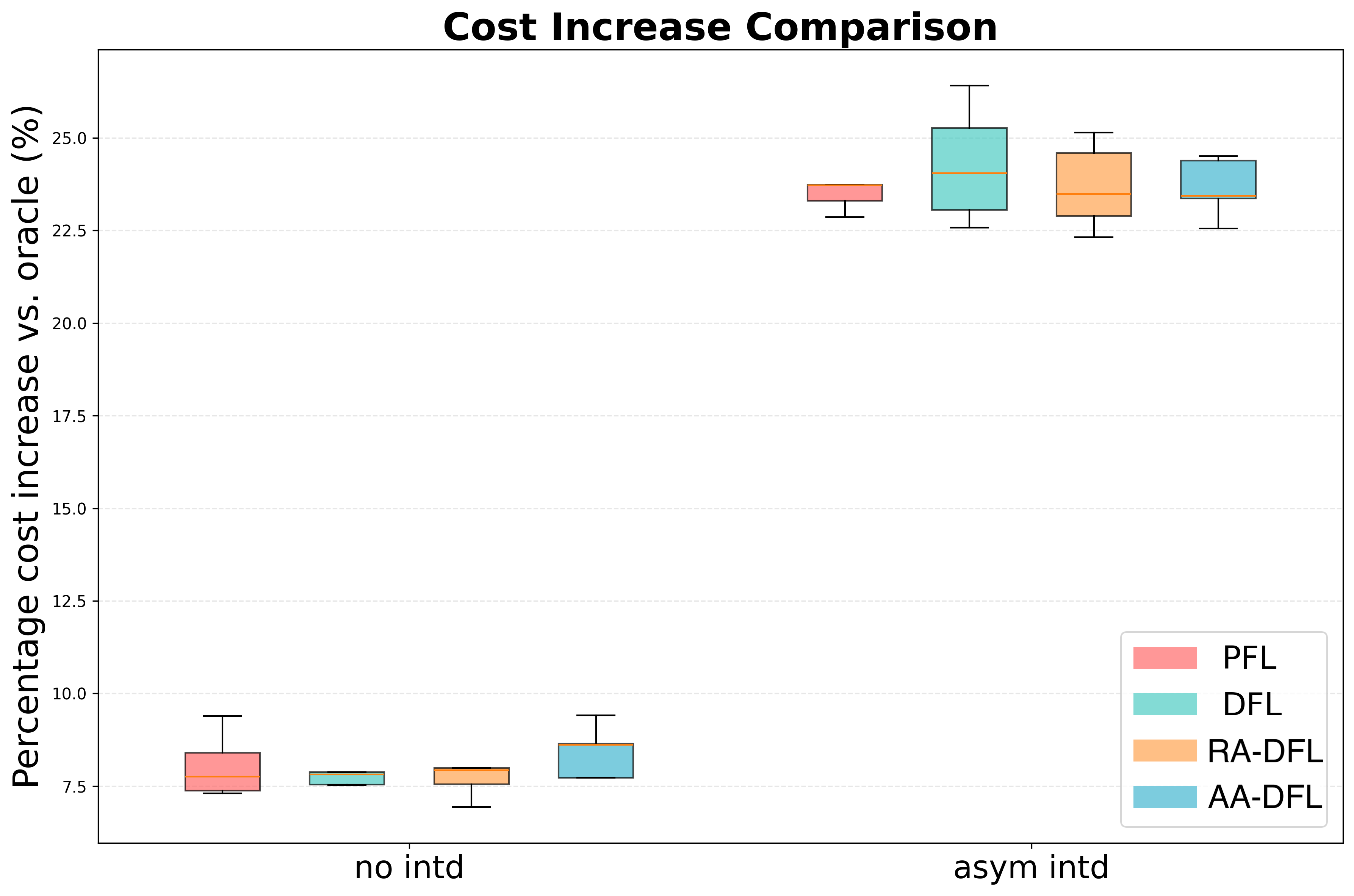}
        \caption{Polynomial degree $deg = 3$.}
        \label{fig:grid_results_deg4}
    \end{subfigure}
    \begin{subfigure}[b]{0.49\textwidth}
        \centering
        \includegraphics[width=\textwidth]{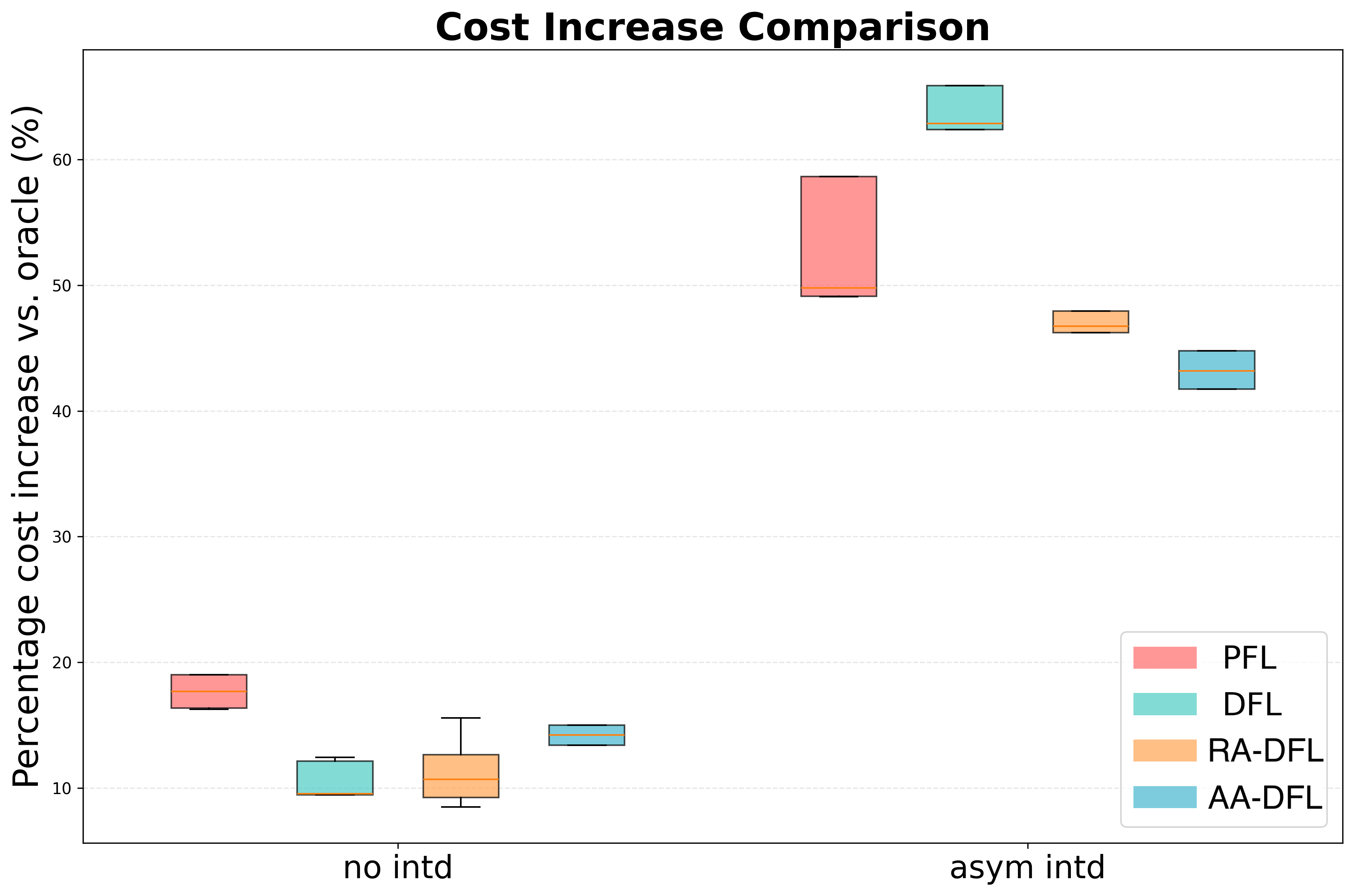}
        \caption{Polynomial degree $deg = 8$.}
        \label{fig:grid_results_deg8}
    \end{subfigure}
    \caption{Grid Graph Results.}
    \label{fig:grid_results}
\end{figure}

Figure~\ref{fig:grid_results} shows the results from the $5\times 5$ grid simulations. It displays $R_{sim}$ over 5 simulations for uninterdicted and interdicted settings. For the low degree of $deg = 3$, we notice that in the uninterdicted case, the PFL average regret is similar to the DFL methods. PFL, however, has a higher upper quartile and maximum regret than DFL and RA-DFL. PFL has the lowest maximum regret and a similar average regret as the DFL methods in the game setting. It stands out that DFL has a significantly wider spread over 5\%, compared to about 1\% for PFL.

For the higher degree, $deg = 8$, all DFL techniques have a maximum below the minimum of PFL. While the average regret of DFL and RA-DFL is similar in the uninterdicted setting, AA-DFL performs about 2\% (20\% relative) worse. When interdicted, however, the DFL performance degrades significantly to an about 13\% (26\% relative) worse than PFL. This trend is reversed for {RA-DFL} and {AA-DFL}, which outperform the PFL average regret by about 3\% and 6\% (6\% and 14\%), respectively.

We attribute the similar performance of PFL and DFL techniques at $deg = {3}$ to the fact that the mapping is simple enough for all models to learn the true function sufficiently well. For a higher degree ($deg = 8$), the model cannot learn the mapping well-enough anymore. In this case, DFL methods can find a better approximation that leads to better decisions than PFL. In accordance with our observations in Section~\ref{sec:toy}, DFL performs significantly worse in the game setting. This trend, however, is reversed when an A-DFL training method is used. The difference between RA-DFL and AA-DFL is small in this setting. We attribute this to the low number of interdictions relative to the number of edges ({$40 / 5 = 8$}), which increases the chances that strategically important edges are covered by the random samples during training. Nonetheless, we observe, to a small extent, a better generalization of RA-DFL in the uninterdicted case and a better performance of AA-DFL in the game.

\begin{figure}[t]
    \begin{subfigure}[b]{0.49\textwidth}
        \centering
        \includegraphics[width=\textwidth]{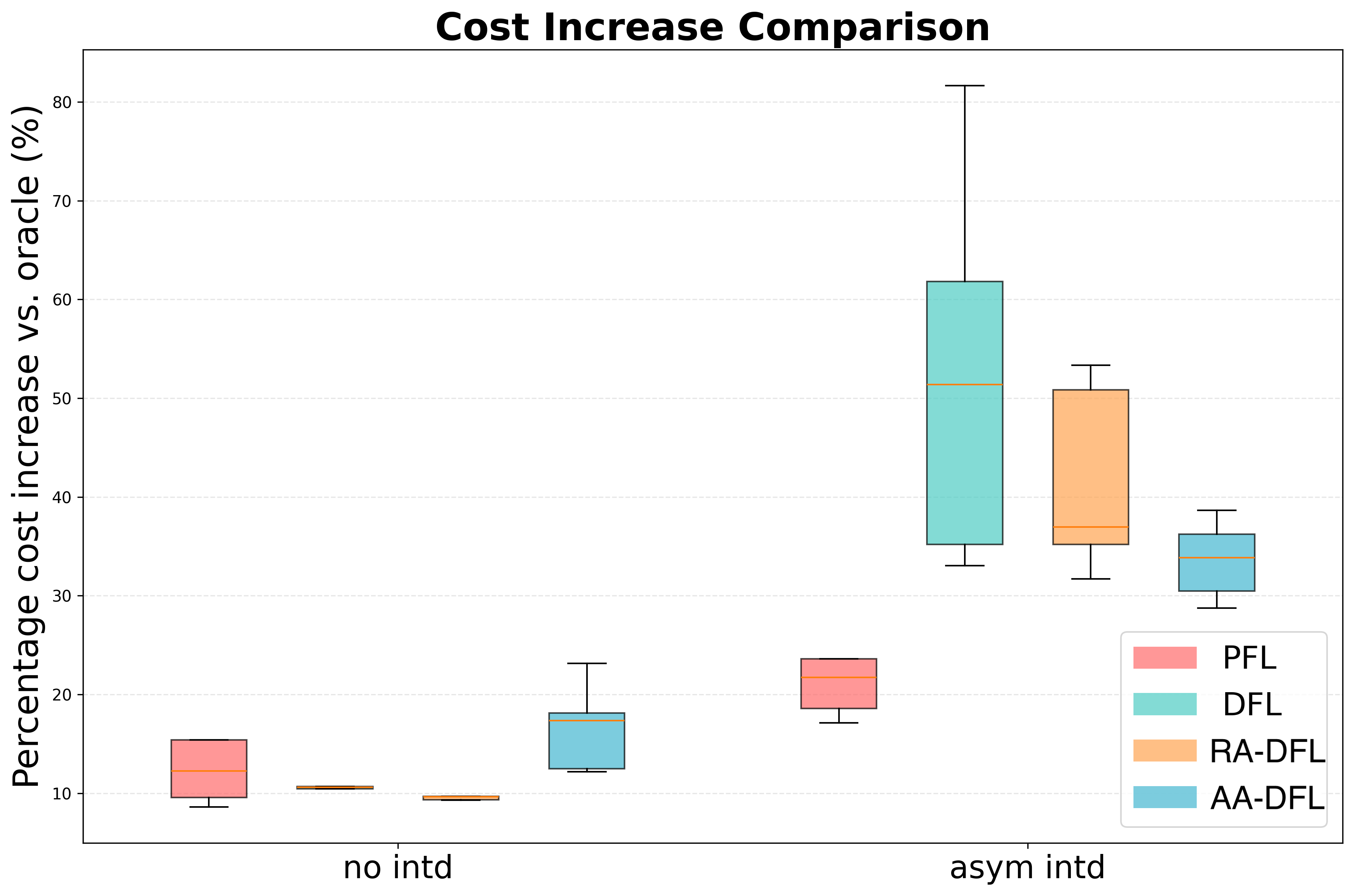}
        \caption{Polynomial degree $deg = 10$.}
        \label{fig:anaheim_results_deg10}
    \end{subfigure}
    \begin{subfigure}[b]{0.49\textwidth}
        \centering
        \includegraphics[width=\textwidth]{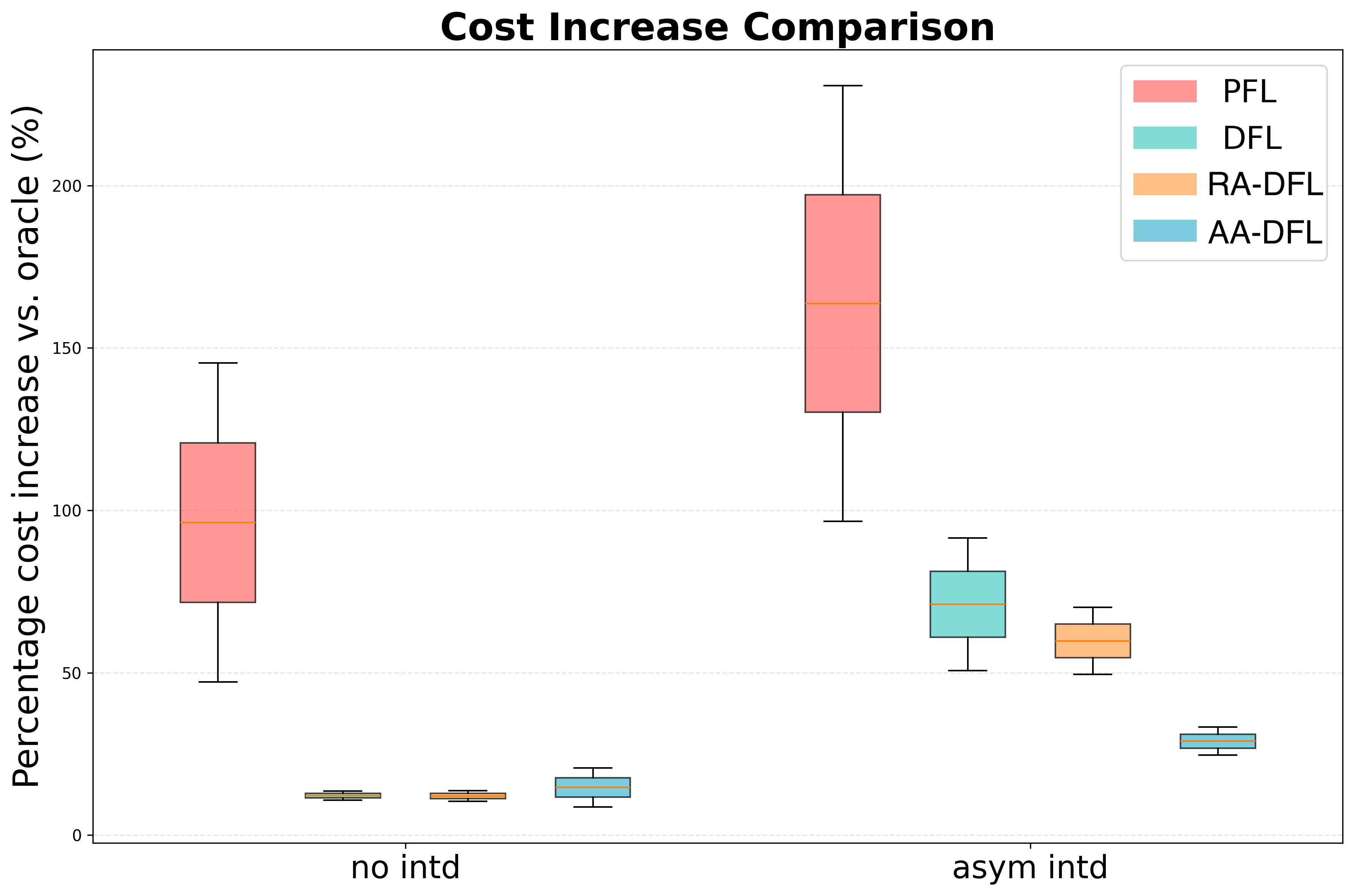}
        \caption{Polynomial degree $deg = 12$.}
        \label{fig:anaheim_results_deg12}
    \end{subfigure}
    \caption{Real-World Graph Results.}
    \label{fig:anaheim_results}
    \vspace{-.7cm}
\end{figure}

The results for the real-world graph topology are depicted in Figure~\ref{fig:anaheim_results}. In this case we used higher degrees ($deg \in \{10, 12\}$), 
to account for the better modeling capabilities of the neural network model.

The lower-degree case with $deg = 10$ shows an intermediate regime. Unlike the lower-degree case in Figure~\ref{fig:grid_results_deg4}, DFL and RA-DFL already outperform PFL in their average regret and show a significantly smaller variance. AA-DFL has an average regret that lies about 6\% above PFL, and 7\% and 8\% above DFL and RA-DFL, respectively. All DFL methods are outperformed by PFL in the game setting, with 
the PFL maximum lying more than 5\% below the minimum of all other methods.

The real-world results show a more pronounced separation between methods for $deg=12$, where all DFL methods outperform PFL significantly with an average regret difference of no less than 70\% (about 350\% relative). This significant performance difference also translates into the game setting, where DFL, RA-DFL, and AA-DFL have an average regret improvement over PFL of 100\%, 110\%, and 150\%, respectively. We also observe that AA-DFL outperforms DFL and RA-DFL by 50\% and 40\% (166\% and 133\% relative).

The real-world example strengthens our observation from the grid example that DFL techniques outperform PFL techniques at higher degrees; i.e., for more complex feature-cost mappings relative to model complexity. 
We also observe a larger performance gap between AA-DFL and RA-DFL. We attribute this to the significantly larger number of edges relative to the number of interdictions: $914 / 5 = 182.8$. Fewer interdictions mean that RA-DFL is less likely to cover strategically important edges during training, giving the focused adversarial selection of AA-DFL an advantage.

%%%%%%%%%%%%%%%%%%%%%%%%%%
%%%%% Conclusion %%%%%%%%%
%%%%%%%%%%%%%%%%%%%%%%%%%%
\section{Conclusion}

We have shown that Decision-Focused Learning (DFL) faces a fundamental structural failure when deployed in Shortest-Path Network Interdiction (SPNI) games. Its training objective admits a wide decision-equivalence class of cost estimators that achieve zero nominal loss yet diverge from the true costs in ways that matter under interdiction. The interdictor can exploit this misalignment to steer the evader onto suboptimal paths, reversing DFL's usual advantage over even an unsophisticated Prediction-Focused Learning (PFL) strategy in the game setting. We characterized this failure mode analytically in a toy example and confirmed it empirically.

We present A-DFL as a remedy. A-DFL works by augmenting nominal training samples with interdicted scenarios, and through this, collapses the harmful equivalence class, supplying a gradient signal in exactly the regime where DFL fails. We introduced two variants: RA-DFL, which selects interdiction scenarios uniformly at random, and AA-DFL, which selects worst-case interdictions using the true costs. We view RA-DFL as the practical default, because it is cheap to implement and effective across settings. AA-DFL, on the other hands, can be viewed as preferable when the graph is large relative to the interdiction budget, where random selection is unlikely to cover the strategically relevant arcs.

Experiments on synthetic grid graphs and a real-world transportation network confirm three key findings: (1) A-DFL restores DFL's advantage in the game setting without sacrificing performance on the standalone shortest-path problem; (2) the relative benefit of DFL-based methods over PFL grows with problem complexity, and (3) AA-DFL offers additional gains over RA-DFL in sparse interdiction settings, consistent with our attributed mechanism.

Several directions remain open for future research. The decision-equivalence class analysis is developed here for linear predictors; extending it to nonlinear models may reveal a different class structure and warrants further study. The assumption that the evader observes interdiction outcomes at test time is also restrictive, and relaxing it would bring the setting closer to fully adversarial deployment. In addition, conceptually, our work can be viewed as developing better strategies for a second-mover defender (e.g., routing over a network in the presence of a first-mover jammer), or alternatively, as providing a better understanding of the limits faced by a second-mover attacker; in the latter case, developing better first-mover strategies remains a direction of future work. {Finally, the experiments rely on synthetic cost distributions and a fixed graph topology; how these results extend to real observed cost data on larger or dynamic networks remains to be studied. Future work should also systematically study the effect of A-DFL's hyperparameters, such as the size of auxiliary dataset $N_{\mathrm{intd}}$, the interdiction budget $B$, and the choice of RA-DFL vs. AA-DFL, on the trade-off between computational cost and performance.}

\vspace{.009in}
{\small \emph{Disclosure of Interests:} The authors have no competing interests to declare that are relevant to the content of this article.}

% ---- Bibliography ----
\clearpage
\bibliographystyle{splncs04}
\bibliography{references}

%%%%%%%%%%%%%%%%%%%%%%%%%%
%%%%% Appendix %%%%%%%%%%%
%%%%%%%%%%%%%%%%%%%%%%%%%%
\appendix

\vspace{-.2in}
\section{SPO$+$ Safeguard for Graphs with Cycles}
\label{app:spo-fix}
\vspace{-.1in}

The SPO$+$ surrogate~\cite{elmachtoub2022spo} approximates the decision-loss subgradient as $\b{y}^\ast(2\hat{\b{c}} - \b{c}) - \b{y}^\ast(\b{c})$, which requires solving an auxiliary shortest-path problem with arc costs $2\hat{\b{c}} - \b{c}$. On a directed grid this is harmless, but on a general directed graph $2\hat{\b{c}}-\b{c}$ creates negative-cost cycles whenever $\hat{c}_i < c_i/2$, making the auxiliary LP unbounded. We apply two training-time safeguards (inactive at inference). First, we clip the cost fed to the surrogate, $\tilde{c}_i = \max\{\hat{c}_i,\, c_i/2 + \varepsilon\}$ with $\varepsilon = 10^{-3}$, so that $2\tilde{c}_i - c_i \ge 2\varepsilon > 0$ and the LP stays bounded. Since clipping freezes the gradient for predictions below $c_i/2+\varepsilon$, we restore learning pressure with a hinge penalty $\mathcal{P}(\hat{\b{c}};\b{c}) = \tfrac{\lambda}{m}\sum_{i=1}^{m}\max\{c_i/2 + \varepsilon - \hat{c}_i,\, 0\}$ ($m=|E|$, weight $\lambda>0$), which vanishes in the safe region $\hat{c}_i \ge c_i/2+\varepsilon$ and grows linearly below it.

\vspace{-.1in}
\section{Toy example details}
\label{app:toy}
\vspace{-.1in}

\noindent\textbf{Initialization:} 
The initialization $a_1 = -a_2 = 0.1$, $\b b = \b 0$ was chosen for visual clarity: it causes A-DFL to converge near the true function, making the improvement over DFL apparent in Figure~\ref{fig:toy-adfl-unintd}. Other initializations yield valid solutions within $\mathcal{F}_{A\text{-}DFL}$ that achieve correct post-interdiction decisions, but with slopes rotated away from the true function, obscuring the visual comparison.
\\

\noindent\textbf{Zero gradients of DFL:} 
The estimated function $\hat f(w) \in \mathcal F_{DFL}$ results in a DFL loss $\ell_{DFL} = \b c^\top (\b y^\ast(\hat f(w)) - \b y^\ast (f(w))) = 0$. Let there be a neighborhood of the slopes $\b a$, with $\epsilon > 0$, where $\epsilon < \frac{|a_1 - a_2|}{2}$, such that $f(w) = (a \pm \epsilon)w + b$. In this neighborhood, we get that $\ell_{DFL} = \b c^\top (\b y^\ast(\hat f(w) \pm \epsilon w) - \b y^\ast (f(w)) \pm \epsilon w) = 0$. The zero-neighborhood of the slopes $\b a$ is a subset of the decision-equivalence class $\mathcal F_{DFL}$, so that the DFL loss remains zero. It follows that the gradients with respect to the slopes $\b a$ vanish.
\\

\noindent\textbf{A simplified interdictor:} 
To keep the toy example in Section~\ref{sec:toy} simple, we require the estimated costs to satisfy
\begin{equation*}
\hat{\b c}_i + \b d_i > \hat{\b c}_j, \quad \text{for } i \neq j \text{ with } \b c_i > \b c_j,
\end{equation*}
ensuring the interdictor always interdicts the path with the lower true cost.
\\

{

\noindent\textbf{DFL equivalence class:} 
In this paragraph we show that in the toy example, $\operatorname{argmin}_{f}\{\ell_{\mathrm{DFL}}(f(w), \b c(w))\} = \mathcal F_{\mathrm{DFL}}$. I.e., $\ell_{\mathrm{DFL}}(f(w), \b{c}(w)) = 0,$ for all $w \neq 0$, if and only if $f \in \mathcal{F}_{\mathrm{DFL}}$.\looseness=-1
\\
\emph{Proof:} Given $\b c(w) = [w, -w]^\top$, it follows from~\eqref{eq:dfl-regret}, that for all $w \neq 0$,
$$
\ell_{\mathrm{DFL}}(f(w), \b{c}(w)) = \b{c}(w)^\top(\b{y}^\ast (f(w)) - \b{y}^\ast (\b{c}(w))) = 0,
$$

if and only if $\b{y}^\ast(f(w)) = \b{y}^\ast(\b{c}(w))$. Since $\b{y}^\ast(\b{\xi}) = \operatorname{argmin}(\b{\xi}^\top \b{y})$ and $\b f(w) = \b a w + \b b$, the DFL loss is zero iff the argmin is preserved, i.e., $a_1 w + b_1 < a_2 w + b_2 \Leftrightarrow w < 0$ for all $w \neq 0$. Rearranging, $(a_1 - a_2)w < b_2 - b_1 \Leftrightarrow w < 0$, which holds for all $w \neq 0$ if and only if $a_1 > a_2$ and $b_1 = b_2$. $\square$
\\

\noindent\textbf{A-DFL equivalence class:} 
Next, we show that  $\operatorname{argmin}_{f}\{\ell_{\mathrm{A-DFL}}(f(w), \b c(w))\} = \mathcal F_{\mathrm{A\text{-}DFL}} \subset \mathcal{F}_{\mathrm{DFL}}$. I.e., $\ell_{\mathrm{A-DFL}}(f(w), \b{c}(w)) = 0,\, \forall w$, if and only if $f \in \mathcal{F}_{\mathrm{A\text{-}DFL}}$.
\\
\emph{Proof:} Given $\b c(w) = [w, -w]^\top$, it follows from~\eqref{eq:dfl-regret}, that
$$
\ell_{\mathrm{A-DFL}}(f(w), \b{c}(w)) = (\b{c}(w)^\top + \b d \odot \b x) (\b{y}^\ast(f(w) + \b d \odot \b x) - \b{y}^\ast (\b{c}(w) + \b d \odot \b x)) = 0,
$$

holds for all $w$ if and only if $\b{y}^\ast(f(w)) = \b{y}^\ast(\b{c}(w))$. Since $\b{y}^\ast(\b{\xi}) = \operatorname{argmin}(\b{\xi}^\top \b{y})$ and $\b f(w) = \b a w + \b b$, the DFL loss is zero iff the argmin is preserved, i.e., $a_1 w + b_1 + d_1 x_1 < a_2 w + b_2 + d_2 x_2 \Leftrightarrow 2w < d(x_2 - x_1)$ for all $w \neq 0$. There follow two cases from the choice of $\b x$.

\emph{Case 1:} $x_1 = 0,\, x_2 = 1$. Then, it follows that $(a_1 - a_2)w < b_2 - b_1 + d \Leftrightarrow 2w < d$. This holds if and only if $a_1 - a_2 = 2$ and $b_1 = b_2$.

\emph{Case 2:} $x_1 = 1,\, x_2 = 0$. Then, it follows that $(a_1 - a_2)w < b_2 - b_1 - d \Leftrightarrow 2w < -d$. This holds if and only if $a_1 - a_2 = 2$ and $b_1 = b_2$.

Both cases lead to the same condition, which is exactly the definition of $\mathcal F_{\mathrm{A\text{-}DFL}}$. $\square$
\\

}

\end{document}